\documentclass[11pt,letterpaper]{article}

\usepackage{caption}
\usepackage[top=0.9in,left=1.4in,bottom=1.1in]{geometry}
\usepackage{amsmath,amsthm,amssymb}
\usepackage{enumerate, xcolor, graphicx}
\usepackage{algorithm}
\usepackage[noend]{algpseudocode}
\usepackage{courier}
\usepackage{tcolorbox}
\usepackage{environ}
\usepackage{tikz}

 \usepackage{mathtools}

 \usepackage[hidelinks]{hyperref}
 
\theoremstyle{theorem}

\theoremstyle{definition}

\newcounter{boxlblcounter}  

\NewEnviron{elaboration}{
\par
\begin{tikzpicture}
\node[rectangle,minimum width=1\textwidth] (m) {\begin{minipage}{1\textwidth}\BODY\end{minipage}};
\draw[dashed] (m.south west) rectangle (m.north east);
\end{tikzpicture}
}

\theoremstyle{definition}

\begin{document}
\title{von Neumann and Newman Pokers with Finite Decks}
\markright{von Neumann and Newman Pokers for Finite Decks}
\author{Tipaluck Krityakierne  \and Thotsaporn Aek Thanatipanonda \and Doron Zeilberger
\date{}
}

\maketitle

\begin{abstract}
John von Neumann studied a simplified version of poker where the ``deck'' consists of {\it infinitely} many cards, in fact, all real numbers between $0$ and $1$.
We harness the power of computation, both numeric and symbolic, to investigate analogs with {\it finitely} many cards. We also study finite analogs
of a simplified poker introduced by D.J. Newman, and conclude with a thorough investigation, fully implemented in Maple, of the three-player game,
doing {\it both} the finite and the infinite versions. This paper is accompanied by  two Maple packages and numerous output files.

\end{abstract}

\section{Prelude}
Welcome to the world of poker, where strategy and probability rule. 
Picture yourself at the poker table, every decision a crucial step toward victory or defeat. 
Poker has intrigued mathematicians for decades as a window into decision-making and game theory. Pioneers like \'Emile Borel, John von Neumann, Harold W. Kuhn, John Nash, and Lloyd Shapley (\cite{B},\cite{NM},\cite{K},\cite{NS})
who believed that real-life scenarios mirror poker with their elements of bluffing and strategic thinking, have simplified the complexities of the game, making it tractable for game theoretic analysis.

\subsection*{von Neumann Poker}

\begin{figure}[h]
    \centering
    \includegraphics[scale=0.4]{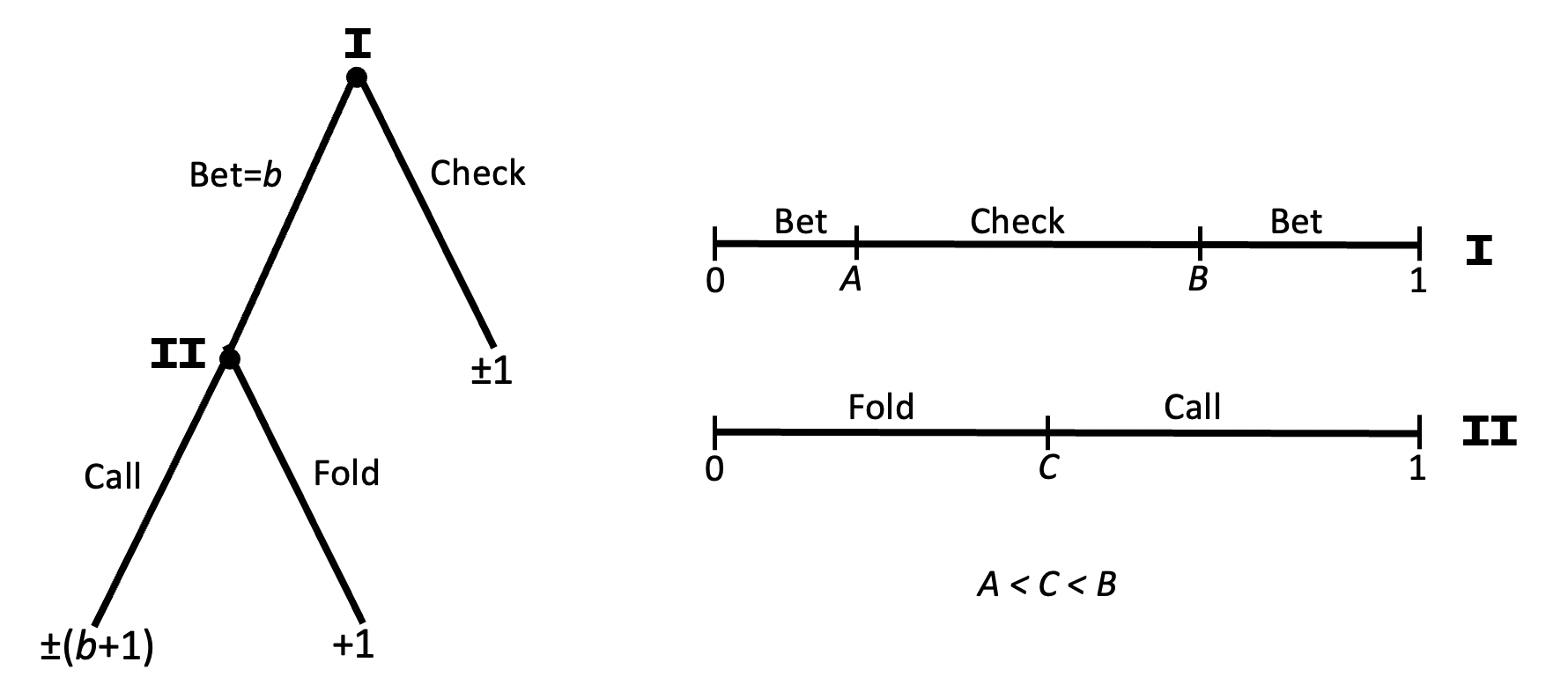}
    \caption{The betting tree and Nash equilibrium strategies for von Neumann Poker}
    \label{fig:Player2_strategies}
\end{figure}

In the original version, von Neumann proposed, and solved, the following game of poker with an uncountably infinite deck, namely all the real numbers between $0$ and $1$. 
Fix a bet size, $b$. Player I and Player II are dealt (uniformly at random) two ``cards'', real numbers $x$ and $y$, in the interval $[0,1]$. They each see their own card, but have no clue about the opponent's card. At the start they each put one dollar into the pot (the so called {\it ante}), so now the pot has two dollars. 

Figure \ref{fig:Player2_strategies} illustrates the ``betting tree'' of this game. Here, Player I looks at his card, and decides whether to {\it check}, in which case each of the players shows their cards, and whoever has the largest card wins the pot. On the other hand he has an option to {\it bet}, putting $b$ additional dollars in the pot. Now the game turns to Player II. She can decide to {\it fold}, in which case player I gets the pot, resulting in a gain of $1$ dollar for Player I, (and a loss of $1$ dollar for player II), or be {\bf brave} and {\it call}, putting
her own $b$ dollars into the pot, that now  has $2b+2$ dollars. The cards are compared in a {\it showdown} and whoever has the
larger card, wins the whole pot, resulting in a gain of $b+1$ dollars for the winner, and a loss of $b+1$ for the loser. 

von Neumann proved that the following pair of strategies is a pure {\it Nash Equilibrium}, i.e. if the players both follow their chosen strategy,
neither of them can do better (on average) by doing a different strategy.

\subsubsection*{The von Neumann advice}

von Neumann identified the cuts $A,B$ and $C$ in the right panel of Figure \ref{fig:Player2_strategies}, and proposed the following strategies.

\begin{itemize}
    \item Player I: If $0<x<\dfrac{b}{\left(b +4\right) \left(b +1\right)}$ or $\dfrac{b^{2}+4 b +2}{\left(b +4\right) \left(b +1\right)}<x<1$ you should {\bf bet}, otherwise {\bf check}.
    \item Player II: If $0<y< \dfrac{b \left(b +3\right)}{\left(b +4\right) \left(b +1\right)}$ you should {\bf fold}, otherwise {\bf call}.
\end{itemize}

Note that Player II's strategy corresponds to {\it honest common sense}, there is some {\it cut-off} that below it you should be conservative, and ``cut your losses''
giving up the one dollar, and not risking losing $b$ additional dollars, and above it, be brave, and {\it go for it}.

Now an {\it honest common sense} would tell you that Player I would also have his own cutoff, check if his card is below it, and
bet if it exceeds it. But this is {\bf not} optimal. If Player I has
a low card, he should {\bf bluff}, and `pretend' that he has a high card, and player II would be intimidated into folding.

{\bf Sad but true}, ``honesty is {\bf not} the best policy''.  Indeed the game favors Player I, and
his expected gain is $\dfrac{b}{\left(b +4\right) \left(b +1\right)}$.

When $b=2$, the advice spells out as follows:
\begin{itemize}
\item Player I: if $0<x<\frac{1}{9}$ or $\frac{7}{9}<x<1$ you should {\bf bet}, otherwise {\bf check}.

\item  Player II: If $0<y< \frac{5}{9}$ you should {\bf fold}, otherwise {\bf call}.
\end{itemize}
The expected value, i.e. the value of the game (for Player I) is $\frac{1}{9}$. It can be shown that $b=2$ maximizes Player I's payoff under the Nash equilibrium strategies.

\subsection*{Finitely Many Cards}

What we {\it don't} like about the original von Neumann version is that the deck is infinite. In {\bf real life} there are only {\it finitely} many cards,
and in fact, not that many. 
We were wondering whether there exists pure Nash equilibria when there are only finitely many cards.

We hope that you would download the  Maple package {\tt FinitePoker.txt}, available, free of charge, from {\small\url{https://sites.math.rutgers.edu/\~zeilberg/tokhniot/FinitePoker.txt}} . Once you downloaded our Maple package to your laptop, that has Maple, and set the directory to be the one where the package resides, start a worksheet
and type {\tt read `FinitePoker.txt`} .

We wrote procedure {\tt vnNE(n,b)}, that inputs 

\begin{itemize}
  \item a positive integer {\tt n}, at least  $2$, standing for the number of cards in the deck, that are numbered $1,2, \dots, n$.
  \item a positive integer {\tt b}, at least $1$, denoting the (fixed) {\it bet size}.
\end{itemize}

It outputs the set of {\bf all} pure Nash equilibria. 
This set may be empty, since we are talking about {\bf pure} NEs (from now on NE:=Nash Equilibrium). Recall that thanks to John Nash, we are only {\bf guaranteed} the existence of {\it mixed} NEs. 
We will talk about mixed NEs later on. But for now, let's explore the pure ones.

\subsubsection*{Finding all pure Nash Equilibria via the ``Vanilla'' approach}

We did not make {\it any} assumptions about `plausible' strategies, so {\it a priori}, a strategy for player I can be {\it any} subset, $S_1$, of $\{1, \dots , n\}$,
that advises: `If your card belongs to $S_1$ you should {\bf bet}, otherwise, {\bf check}'. Similarly a strategy for player II, $S_2$, can be any such subset,
that tells her to call iff her card $j \in S_2$. For each conceivable strategy pair $[S_1,S_2]$ we can easily compute the expected payoff following these strategies.
This is implemented in procedure
{\tt EnS1S2(n,S1,S2,b)} .

Using this, we can construct the {\it paytable}, implemented in procedure  {\tt PayTable(n,b)}, that is a $2^n$ by $2^n$ matrix. Now we look
for pure NEs, the usual way, by finding, for {\it each} strategy of each player the {\it best response} of the other player, and looking for pairs $[S_1,S_2]$ that are best responses to each other.

Let's fix the bet size $b=2$. If the card has only $2$  cards, {\tt vnNE(2,2);} gives
$$
\{[\{\}, \{2\}, 0], [\{2\}, \{2\}, 0]\} \quad,
$$
so there are two pure NEs. In both of them Player II bets if her card is $2$ and folds if her card is $1$, while
Player I always checks in the first strategy, and checks if his card is $1$ in the second strategy.

This is not very interesting, since the expected gain (value of the the game) is $0$.

{\tt vnNE(3,2)} is equally boring, giving the two trivial pairs $[\phi,\{3\}]$ and $[\{3\},\{3\}]$ .

{\tt vnNE(4,2)},  {\tt vnNE(5,2)}, and {\tt vnNE(6,2)} are even more boring, they are empty!

But now comes a nice surprise, {\tt vnNE(7,2)} gives three pure, {\it non-trivial},  NEs. 

For all of them Player I  bets iff his card belongs to $\{1,6,7\}$, but
Player II calls if her card is in either $\{3,6,7\}$, $\{4,6,7\}$, or $\{5,6,7\}$. The value of the game is $\frac{2}{21}$.

So with $7$ cards we already have bluffing! If Player I has the card labeled $1$, he should bet even though he would
definitely lose the bet if Player II calls. 

Moving right along, {\tt vnNE(8,2);} also gives you three pure NEs.

For all of them Player I  bets iff his card belongs to $\{1,7,8 \}$, but
Player II calls if her card is in either $\{4,7,8\}$, $\{5,7,8\}$, or  $\{6,7,8\}$. The value of the game is $\frac{3}{28}$, getting 
tantalizingly close to von Neumann's $\frac{1}{9}$.

Since the sizes of the payoff matrices grow exponentially, and we did not make {\it any} {\it plausibility assumptions}, there is only so far we can
go with this naive {\it vanilla} approach. But nine cards are still doable. Indeed there are seven pure NEs in this case.
For all of them $S_1=\{1,8,9\}$, but Player II has seven choices, all with four members, including, of course, $\{6,7,8,9\}$.

For all pure NEs for $n$ from $2$ to $10$ and bet-sizes from $1$ to $5$ look at the output file:
{\small\url{https://sites.math.rutgers.edu/~zeilberg/tokhniot/oFinitePoker1.txt}}.

To overcome the {\it exponential explosion}, we can stipulate that Player I's strategy {\bf must} be of the form:

``Check iff $i \in \{A,A+1, \dots,B\}$ for some $1\leq A<B \leq n$, ''

while Player II's must be of the form:

``Call iff $j \in \{C,C+1, \dots,n\}$ for some $1\leq C  \leq n$.''

Now we can go much further, see the output file at \\
{\small\url{https://sites.math.rutgers.edu/~zeilberg/tokhniot/oFinitePoker1A.txt}} \\
for pure NEs for $n$ up to 27.

If $n$ is a multiple of $9$ then the (restricted) pure NEs are as expected, namely  the value of the game is $\frac{1}{9}$ and
the strategy for player I is: check if $\frac{1}{9}n<i\leq \frac{7}{9}n$, bet otherwise
and for Player II: call iff $j>\frac{5}{9}n$.

If $n$ is not a multiple of $9$, then the values are close, but a little less.
For example for $n=26$ the value is $\frac{36}{325}=0.110769$. For $n=25$ the value is
$\frac{11}{100}=0.11$, for $n=24$ it is $\frac{61}{552}= 0.1105072464$, for $n=23$ it is
$\frac{28}{253}= 0.1106719368$, for $n=22$ it is $\frac{17}{154}=0.1103896104 $.

\section{Mixed NEs via Linear Programming}

The study of mixed strategies in two-person zero-sum games can be elegantly formulated as a primal-dual linear programming (LP) problem. A {\it mixed} strategy involves each player choosing optimal actions according to a probability distribution, introducing uncertainty. 
An equilibrium solution to this dual pair of linear programs reveals optimal mixed strategies (mixed NE) for both players.

\subsubsection*{Slow LPs for mixed NE}

Recall our scenario: the pot starts at 1+1, with only Player I able to bet a fixed amount $b$. Given the $2^n$ by $2^n$ payoff matrix $\left(m_{ij}\right)$ as input, Player I aims to maximize his worst-case expected gain, minimizing over all possible actions of Player II. This objective is framed as an LP by introducing variable $v_1$ to represent this minimum, ensuring Player I's expected gain is at least  $v_1$ for every action of Player II, and maximizing $v_1$. 
Similarly, from Player II's viewpoint, the goal is to minimize her worst-case expected loss, maximizing over all actions of Player I. This involves introducing variable $v_2$ to represent this maximum, and setting the objective to minimize $v_2$. 

To formulate the primal-dual LP,  let ${\bf{x}}=\left(x_1,\dots,x_{2^n}\right)$ be the mixed strategy probability of Player I to maximize $v_1$. Let ${\bf{y}}=\left(y_1,\dots,y_{2^n}\right)$ be the mixed strategy probability of Player II to minimize $v_2$.

\begin{alignat*}{2}
    & \begin{aligned} 
 & \text{Primal: Maximize } v_1 \\
\text{s.t. } & \sum_{i=1}^{2^n} x_i \cdot m_{ij} \geq v_1\quad \text{for } j = 1, ..., 2^n \\
& \sum_{i=1}^{2^n}x_i =1 \\
&x_i \geq 0 \quad \text{for } i = 1, ..., 2^n. \\
  \end{aligned}
    & \hskip 6em &
  \begin{aligned}
 & \text{Dual: Minimize }  v_2 \\
\text{s.t.  } & \sum_{j=1}^{2^n}   m_{ij}\cdot y_j \leq v_2\quad \text{for } i = 1, ..., 2^n \\
& \sum_{j=1}^{2^n}y_j =1 \\
&y_j \geq 0 \quad \text{for } j = 1, ..., 2^n. \\
  \end{aligned}
\end{alignat*}

\vspace{2em}

By the minimax theorem at an equilibrium, $v_1=v_2=v^*$, which represents the value of the game.
We can use the commands {\tt maximize} and {\tt minimize} in the Maple package {\tt simplex}, to find
{\bf one} mixed NE. 
However, due to the exponentially large size of the matrix, practical limitations arise, restricting us from considering more than 6-7 cards without the inconvenience of reducing the dominated rows/columns of the payoff matrix (worse than the vanilla approach in the previous section).

We have implemented the above Slow LP procedures in {\tt MNE(M,S1,S2)}. Typing \\ 
\texttt{M := PayTable(4,2)[1]:\\
S1 := Stra1(4):\\
S2 := Stra2(4):\\
MNE(M,S1,S2); \quad
}
gives outputs:\\
\noindent \texttt{ [\{[\{4\}, 1/2], [\{1,4\}, 1/2]\}, \{[\{4\}, 1/2], [\{2,4\}, 1/2]\}, 1/12] }.

\vspace{1em}
Translation:
\begin{itemize}
    \item The value of the game is $\frac{1}{12}$ (last entry).
    \item Player I has two strategies specified within the first set of braces: (1.1) with probability 1/2, raise if his card is 4 and fold if his cards are 1, 2, or 3; and (1.2) with probability 1/2, raise if his card is 1 or 4 and fold if his cards are 2 or 3.
    \item Player II  has two strategies specified within the second set of braces: (2.1) with probability 1/2, call if her card is 4 and fold if her cards are 1, 2, or 3; and (2.2) with probability 1/2, call if her cards are 2 or 4 and fold if her cards are 1 or 3.
\end{itemize} 

\subsubsection*{Fast LPs for mixed NE}

The NE can be considered from another point of view, whereby focusing on the card each player receives can reduce the number of constraints from exponential to linear. With this formulation, we can handle more than 200 cards now.

A strategy for Player I is given by a vector ${\cal{P}}=[p_1, \dots, p_n]$ that tells him: if his card is $i$, bet with probability $p_i$, and check with probability $1-p_i$. 

A strategy for Player II is given by a vector ${\cal{Q}}=[q_1, \dots, q_n]$ that tells her: if her card is $j$, call  with probability $q_j$, and fold with probability $1-q_j$.

Before we discuss the Fast LP formulation, let's mention that given card-by-card strategies, $\cal{P}$ and $\cal{Q}$, it is easy to compute the {\it expected payoff} (for Player I). This is implemented in procedure {\tt PayOffP1P2(n,b,P1,P2)}, as a bilinear form in the $p_i$'s and $q_j$'s:

\begin{align*}
 &   \frac{1}{n(n-1)}
\Bigg( \Bigg.\sum_{i=1}^{n} \sum_{j=1}^{i-1} (1-p_i) \, 
-\sum_{i=1}^{n} \sum_{j=i+1}^{n} (1-p_i) \,
+\sum_{i=1}^{n} \sum_{j=1}^{i-1} p_i (1-q_j) \\
&+\sum_{i=1}^{n} \sum_{j=i+1}^{n} p_i (1-q_j) \,
+(b+1)  \sum_{i=1}^{n} \sum_{j=1}^{i-1} p_i q_j \,
-(b+1)  \sum_{i=1}^{n} \sum_{j=i+1}^{n} p_i q_j\Bigg)  .
\end{align*}

Let's now get back to the Fast LP for Player I, which contains two sets of constraints. Each set corresponds to the expected payoff  (over distribution $\cal{P})$,  conditioned on the card that Player II has and whether she calls or folds:
\begin{align*}
\label{eq:VN-I}
& \text{Maximize } \frac{1}{n}\sum_{j=1}^n v_j   \\
\text{s.t. } &
\frac{1}{n-1}\sum_{i\neq j} \left(\texttt{Call}(i,j,b+1)\cdot p_i + \texttt{Call}(i,j,1)\cdot (1-p_i)\right)
\geq v_j \quad j=1,\dots, n \text{ (Player II calls)}\\
&\frac{1}{n-1}\sum_{i\neq j} \left( p_i + \texttt{Call}(i,j,1)\cdot (1-p_i)\right)  \geq v_j \quad j=1,\dots, n \text{ (Player II folds)}\\
& 0\leq p_i \leq   1 \quad i=1,\dots, n, \tag{VN-I}
\end{align*}
where the procedure $\texttt{Call}(i,j,R)$ is defined based on whether the card $i$ is larger than card $j$ or not:
\[\texttt{Call}(i,j,R) = \begin{cases} 
R & \text{if } i > j \\
-R & \text{if } i < j. 
\end{cases}\]

Similarly, for the Fast LP for Player II, the constraints are calculated based on the expected loss (over distribution $\cal{Q})$, conditioned on the card that Player I has and whether he raises or checks:
\begin{align*}
\label{eq:VN-II}
& \text{Minimize } \frac{1}{n}\sum_{i=1}^n v_i   \\
\text{s.t. } & 
\frac{1}{n-1}\sum_{j\neq i} \left(\texttt{Call}(i,j,b+1)\cdot q_j + (1-q_j) \right) \leq v_i \quad i=1,\dots, n \text{ (Player I raises)}\\
& \frac{1}{n-1}\sum_{j\neq i} \texttt{Call}(i,j,1)
\leq v_i \quad i=1,\dots, n \text{ (Player I checks)}\\
& 0\leq q_j \leq 1 \quad j=1,\dots, n. \tag{VN-II}
\end{align*}

The procedure for the Fast LPs is {\tt vnMNE(n,b)}. Now things get interesting sooner. Already with three cards, we have bluffing! 

With bet size $1$, typing {\tt lprint(vnMNE(3,1));} outputs:

{\tt [1/18, .5555555556e-1, [1/3, 0, 1], [0, 1/3, 1]]} \quad .

\vspace{1em}
Translation: 
\begin{itemize}
    \item The value of the game is $\frac{1}{18}$.
    \item  Its value in decimals is $0.055555\dots$.
    \item Player I's strategy is: If your card is $1$, bet with probability $\frac{1}{3}$ and check with probability $\frac{2}{3}$. If your card is $2$ then {\bf definitely check}, while if your card is $3$ then you should {\bf definitely bet}.
    \item Player II's strategy is: If your card is $1$, {\bf definitely fold}, if your card is $2$, call with probability $\frac{1}{3}$ and fold with probability $\frac{2}{3}$, while if your card is $3$ then {\bf definitely call}.
\end{itemize}

So already with three cards, Player I should sometimes bluff if his card is $1$, but only with probability $\frac{1}{3}$.

The output file {\small\url{https://sites.math.rutgers.edu/\~zeilberg/tokhniot/oFinitePoker3.txt}} contains one mixed NE for each of the cases $n$ (size of the deck) from $2$ to $40$, and $b$ (size of the bet) from $1$ to $10$.

The verbose form of {\tt vnMNE(n,b);} is {\tt vnMNEv(n,b);}, spelling out the advice.

Note that a pure NE is also a mixed one, and indeed sometimes we get pure NEs. For example,

{\tt lprint(vnMNE(9,2));} gives:

{\tt [1/9, .1111111111, [1, 0, 0, 0, 0, 0, 0, 1, 1], [0, 0, 0, 0, 0, 1, 1, 1, 1]]}.

\vspace{1em}
Translation: 
\begin{itemize}
\item The value of the game is $\frac{1}{9}$.
\item The value of the game in floating-point is $0.111111111\dots$.
\item Player I: Bet iff your card is in $\{1,8,9\}$. 
\item Player II: Call iff your card is in $\{6,7,8,9\}$. 
\end{itemize}

This was {\it so much faster} than {\tt vnNE(9,2)}.

While {\tt vnNE(18,2)} would take centuries (and run out of memory), {\tt lprint(vnMNE(18,2));} gives  you right away:

{\tt [1/9, .1111111111, [1, 1, 0, 0, 0, 0, 0, 0, 0, 0, 0, 0, 0, 0, 1, 1, 1, 1], [0,0, 0, 0, 0, 0, 0, 0, 0, 0, 1, 1, 1, 1, 1, 1, 1, 1]]}.

Again a pure NE, exactly as in the von Neumann model.

One more example before we move on to DJ Newman's poker.

With $28$ cards and bet size $4$, {\tt lprint(vnMNE(28,4));} gives:

{\tt
[113/1134, .9964726631e-1, [1, 1, 2/3, 0, 0, 0, 0, 0, 0, 0, 0, 0, 0, 0, 0, 0, 0
, 0, 0, 0, 0, 0, 0, 0, 1, 1, 1, 1], [0, 0, 0, 0, 0, 0, 0, 0, 0, 0, 0, 0, 0, 0,
0, 0, 0, 0, 0, 1/3, 1, 1, 1, 1, 1, 1, 1, 1]]}.

We will let you, dear reader, do the translation.

\section{DJ Newman Poker}

Not as famous as John von Neumann, but at least as brilliant, is Donald J. Newman, the third person to be Putnam fellow in three consecutive years.
He was a good friend of John Nash. In a fascinating four-page paper \cite{N} in {\it Operations Research}, he proposed his own version
of poker, where the bet size is {\bf not} fixed, but can be decided by Player I, including  betting $0$, that is the same as {\it checking}.

In his own words (now the players are $A$ and $B$) :

{\it
\quad A and B each ante $1$ dollar and are each dealt a `hand,' namely a randomly chosen real number in $(0,1)$. Each sees his, but not the opponent's hand.
\quad A bets any amount he chooses ($\geq 0$);
\quad B `sees' him (i.e. calls, betting the same amount) or folds.
\quad The payoff is as usual.
}

But in real life, there is always a finite number of cards, and no one can bet arbitrarily large amounts. Once again, we focus on the finite deck version, which is set up as follows: The inputs are integers $n\geq 2$ and $b\geq 1$, where each player is dealt a different card from $\{1, \dots, n\}$, and Player I's decision, upon seeing his card $i$, is to choose an amount $s$ from $\{0, \dots, b\}$ to bet, where $s=0$ corresponds to checking. 

In this game the number of strategies are even larger, and we will not bother with the `vanilla' approach to find pure NEs.
Instead, we will look for (Fast LP) mixed strategies right away.

\subsection*{Player I's payoff maximization} 
Player I's strategy space consists of $n \times (b+1)$ matrix, $\left(p_i[s]\right)$, where $p_i[s]$  ($1 \leq i \leq n, 0 \leq s \leq b$)
is the probability that if he has card $i$, he would bet $s$ dollars (of course, the row-sums should add-up to $1$). 
The LP formulation is analogous to that of \eqref{eq:VN-I} in the previous section. Let's point out the differences to gain some insights. Recall that each constraint corresponds to the card that Player II has and her choice of action. In \eqref{eq:VN-I}, Player II can either call or fold, and she can have one of the $n$ cards. Hence, there are a total of $2n$ constraints.

In our current scenario, however, Player II's decision depends on both her card and Player I's proposed bet amount $s$. Let  $S_b:=\{0,\dots,b\}$. We define ${\cal{P}}(S_b)$ as the set containing all possible strategies of Player II regarding whether to call or fold.
That is, each $Y\in {\cal P}(S_b)$ represents a strategy where Player II will call if $s\in Y$. Therefore, for a fixed card $j$ and strategy $Y\in {\cal P}(S_b)$ of Player II, the constraint is: \textit{``Player II calls if she holds card $j$ and the proposed $s\in Y$; otherwise, she folds.''} The total number of these constraints amounts to $n\cdot2^b$.

For example, when $b=4$, 
\begin{align*}
    {\cal{P}}(S_4)=&\{ \{0\}, \{0, 1\}, \{0, 2\}, \{0, 3\}, \{0, 4\}, \{0, 1, 2\}, \{0, 1, 3\}, \{0, 1, 4\}, \{0, 2, 3\}, \{0, 2, 4\} \\
  &  \{0, 3, 4\}, \{0, 1, 2, 3\}, \{0, 1, 2, 4\}, \{0, 1, 3, 4\}, \{0, 2, 3, 4\}, \{0, 1, 2, 3, 4\} \}.
\end{align*}

With this setup, we derive the following LP:
 
\begin{align*}
& \text{Maximize } \frac{1}{n}\sum_{j=1}^n v_j   \\
\text{s.t. } &
\frac{1}{n-1}\sum_{i \neq j}   \left(\sum_{s \in Y} \texttt{Call}(i,j,s+1) \cdot p_i[s]  + \sum_{s \in (S_b \setminus Y)} p_i[s]\right)
\geq v_j \quad \underbrace{j=1,\dots, n;\,\, Y\in {\cal{P}}(S_b)}_{\text{total } n\cdot 2^b \text{ constraints}}\\
& \sum_{s=0}^{b} p_i[s] = 1,\quad i=1,\dots, n\\
&  p_i[s] \geq   0, \quad s=0, \dots, b; \, i=1,\dots, n \tag{DJN-I}
\end{align*}

\subsection*{Player II's loss minimization} 
While Player's II's strategy is also an  $n \times (b+1)$ matrix,  formulating the LP is much simpler. Let's denote the matrix by $\left(q_j[s]\right)$ where $q_j[s]$ is the probability of calling if her card is $j$ and the bet proposed by Player I is $s$ (and as usual $1-q_j[s]$ is the corresponding probability of folding). 
In this case, there are a total of $n(b+1)$ constraints (not exponential as in the case of Player I). Also, the LP formulation straightforwardly extends from \eqref{eq:VN-II}:
\begin{align*}
& \text{Minimize } \frac{1}{n}\sum_{i=1}^n v_i   \\
\text{s.t. } & 
\frac{1}{n-1}\sum_{j \neq i} \left( \texttt{Call}(i,j,s+1) \cdot q_j[s] + (1 - q_j[s]) \right) \leq v_i \quad \underbrace{s=0, \dots, b;\, i=1,\dots, n}_{\text{total } n(b+1) \text{ constraints}} \\
& q_j[0] = 1 \quad j=1, \dots, n \\
& 0\leq q_j[s] \leq 1 \quad s=0, \dots, b;\, j=1,\dots, n. \tag{DJN-II}
\end{align*}

These LPs are implemented in procedure {\tt djnMNE(n,b)}; and the verbose version is {\tt djnMNEv(n,b);}.

We noticed that for any given $n$, there exists a maximal bet size after which the game has the same value. The output file \\{\small\url{https://sites.math.rutgers.edu/\~zeilberg/tokhniot/oFinitePoker4a.txt}} \\
contains one mixed NE for $1\leq n \leq 14$ and for all $b$ until it `saturates'. As $n$ grows larger, and $b$ reaches its saturation value,
the value of the game seems to converge to the DJ Newman `continuous' value $\frac{1}{7}$.

\section{Three-player Poker Game}

As early as 1950, future Economics Nobelists, John Nash and Lloyd Shapley \cite{NS}, pioneered the analysis of a three-player poker game.
They explored a simplified version where the deck contains only two kinds of cards, High and Low, in equal numbers.
However, today, eighty years after von Neumann's 1944 analysis of poker, the dynamics of the three-player game therein remain unexplored. We now take the opportunity to analyze these dynamics in both their finite and infinite versions.

\subsection*{Finite deck}
\begin{figure}
    \centering
    \includegraphics[scale=0.4]{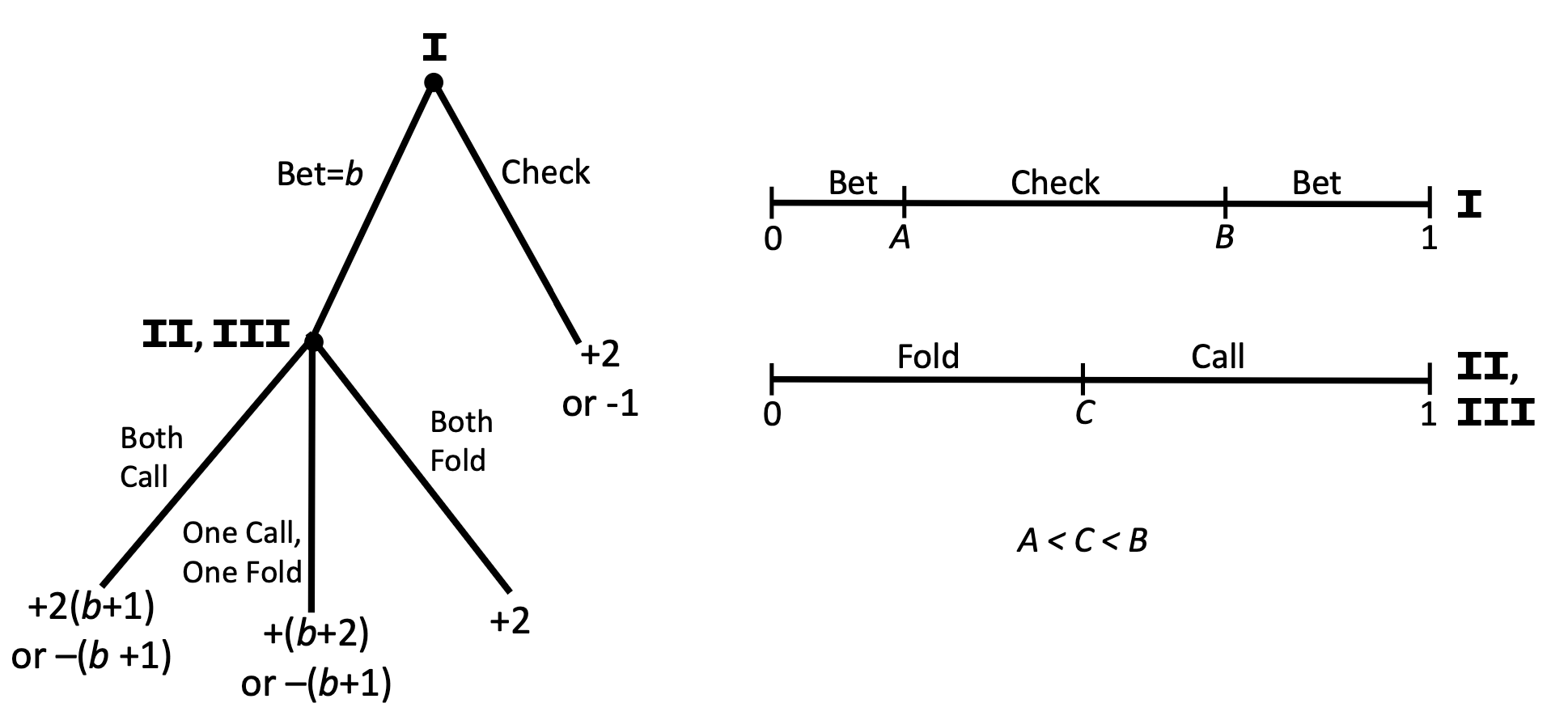}
    \caption{Three-player poker. Left: The betting tree. Right: Conjectured Nash equilibrium strategies for a continuous deck.}
    \label{fig:Player3_strategies}
\end{figure}

The three players each put 1 dollar into the pot. 
Player I acts first, choosing either to check or to bet a fixed integer amount $b > 0$. If Player I checks, the three hands are immediately compared, and the player with the highest hand wins the pot. However, if Player I bets, Players II and III have two choices: call or fold. The reader is invited to refer to the left panel of Figure \ref{fig:Player3_strategies}, which depicts the betting tree for three players. (The right panel shows the conjectured Nash equilibrium strategies to be used in the next section for the continuous version of the game.)

Assume we are given three-dimensional payoff matrices $\left(M_l, l=1, 2, 3\right)$ for the three players: 
\[
M_l = \left(m^l_{ijk}\right), 
\]
where $i, j, k=1,2,\dots,2^n$. 

While its counterpart two-player game can be solved using linear programming, here we require nonlinear programming (NLP) \cite{E}.
The NLP formulation for the three-player game closely follows the LP model for the two players discussed in the previous section. Each player aims to minimize their expected loss, or the expected gain of the other players. For instance, given Player I's payoff matrix  $M_1$, the other two players attempt to minimize the maximum potential loss incurred due to Player I's choices. This involves constraints that utilize matrix $M_1$ and the probability distributions  ${\bf{y}}=\left(y_1,\dots,y_{2^n}\right)$ and ${\bf{z}}=\left(z_1,\dots,z_{2^n}\right)$ of Players II and III. These are embedded in the first set of constraints in the NLP formulation, which we will now formulate. 

The Slow NLP for three players is given by:
\begin{align*}
 & \text{Minimize }  \sum_{l=1}^3 v^l \\
\text{s.t.  } & \sum_{j,k=1}^{2^n}   m^1_{ijk}\cdot y_j \cdot z_k \leq v^1\quad \text{for } i = 1, 2, ..., 2^n \\
& \sum_{i,k=1}^{2^n}   m^2_{ijk}\cdot x_i \cdot z_k \leq v^2\quad \text{for } j = 1, 2, ..., 2^n \\
& \sum_{i,j=1}^{2^n}   m^3_{ijk}\cdot x_i \cdot y_j \leq v^3\quad \text{for } k = 1, 2, ..., 2^n \\
& \sum_{i=1}^{2^n}x_i =1, \qquad \sum_{j=1}^{2^n}y_j =1, \qquad \sum_{k=1}^{2^n}z_k =1 \\
&x_i, y_j, z_k \geq 0 \quad \text{for } i,j,k = 1, 2, ..., 2^n. \\ 
\end{align*}

Note that if there are only two players, $z_k$ in the above NLP formulation disappears, and the constraint functions become linear in the variables $x_i$ and $y_j$. Thus, the problem can be decomposed into two separate LP (primal-dual) problems, as discussed earlier.

Programs related to three players can be found at {\small\url{https://sites.math.rutgers.edu/~zeilberg/tokhniot/ThreePersonPoker.txt}}.
The Slow NLP is implemented in the procedure {\tt MNE(n,b)}, which is, of course, very slow even for small $n$.

We now shift our focus to the Fast NLP formulation for three players, which aligns with the Fast LP formulation for two players, considering on the card each player receives. Recall a strategy for Player I is given by a vector ${\cal{P}}=[p_1, \dots, p_n]$, indicating that if his card is  $i$, he bets with probability $p_i$, and checks with probability $1-p_i$. A strategy for Player II is given by a vector ${\cal{Q}}=[q_1, \dots, q_n]$, indicating that if her card is $j$, she calls  with probability $q_j$, and folds with probability $1-q_j$. Similarly, a strategy for Player III is represented by a vector ${\cal{R}}=[r_1, \dots, r_n]$, following the same interpretation as Player II.

We first define two procedures:
\begin{itemize}
    \item \texttt{Call2} is used to calculate the payoff if either Player II or Player III decides to fold, leaving only two players (one of whom is Player I) to compare their cards. Let us assume that Player III folds. Then, 
\[\texttt{Call2}(i,j,R) = \begin{cases} 
R+1 & \text{if } i > j \\
-R & \text{if } i < j. 
\end{cases}\]

\item \texttt{Call3} is used to calculate the payoff when all the three players are comparing their cards:
\[\texttt{Call3}(i,j,k,R) = \begin{cases} 
2R & \text{if } i > j \text{ and } i>k \\
-R & \text{if } i < j \text{ or } i<k. 
\end{cases}\]
\end{itemize}

The Fast NLP contains three sets of constraints, one set for each player, corresponding to the expected payoff over the pairs of distributions $\cal{Q-R}$, $\cal{P-R}$, or $\cal{P-Q}$. For each player $l=1, 2, 3$, there are two sets of constraints depending on the card that Player  $l$ has and whether they follow their first strategy or the second strategy:

\[
\text{Minimize } \frac{1}{n}\sum_{c=1}^n v^1_c + \frac{1}{n}\sum_{c=1}^n v^2_c + \frac{1}{n}\sum_{c=1}^n v^2_c 
\]
subject to
\begin{align*} 
&\frac{1}{(n-1)(n-2)}\sum_{j\neq i}\sum_{k\neq i,j}  \texttt{Call3}(i,j,k,1)\leq v^1_i \quad i=1,\dots, n \tag{Player I checks}
\end{align*}
\begin{align*} 
& \frac{1}{(n-1)(n-2)}  \Bigg( \Bigg.\sum_{j\neq i}\sum_{k\neq i,j} \texttt{Call3}(i,j,k,b+1)\cdot q_k\cdot r_k \\
&+ \texttt{Call2}(i,j,b+1)\cdot q_j\cdot(1-r_k)
+ \texttt{Call2}(i,k,b+1)\cdot(1-q_j)\cdot r_k \\
&+2(1-q_j)\cdot(1-r_k)\Bigg. \Bigg)
\leq v^1_i \quad i=1,\dots, n \tag{Player I bets}
\end{align*}
\begin{align*}
&\frac{1}{(n-1)(n-2)}\sum_{i\neq j}\sum_{k\neq i,j}\left(
-p_i + \texttt{Call3}(j,i,k,1)\cdot (1-p_i)\right) \leq v^2_j \quad j=1,\dots, n \tag{Player II folds}
\end{align*}
\begin{align*} 
& \frac{1}{(n-1)(n-2)}  \Bigg( \Bigg.\sum_{i\neq j}\sum_{k\neq i,j} \texttt{Call3}(j,i,k,b+1)\cdot p_i\cdot r_k \\
&+\texttt{Call2}(j,i,b+1)\cdot p_i \cdot(1-r_k)
+\texttt{Call3}(j,i,k,1)\cdot(1-p_i) \Bigg. \Bigg)
\leq v^2_j \quad j=1,\dots, n \tag{Player II calls}
\end{align*}
\begin{align*}
&\frac{1}{(n-1)(n-2)}\sum_{i\neq k}\sum_{j\neq i,k}\left(
-p_i + \texttt{Call3}(k,i,j,1)\cdot (1-p_i)\right) \leq v^3_k \quad k=1,\dots, n \tag{Player III folds}
\end{align*}
\begin{align*} 
& \frac{1}{(n-1)(n-2)}  \Bigg( \Bigg.\sum_{i\neq k}\sum_{j\neq i,k} \texttt{Call3}(k,i,j,b+1)\cdot p_i\cdot q_j \\
&+\texttt{Call2}(k,i,b+1)\cdot p_i \cdot(1-q_j)
+\texttt{Call3}(k,i,j,1)\cdot(1-p_i) \Bigg. \Bigg)
\leq v^3_k \quad k=1,\dots, n \tag{Player III calls}
\end{align*}
\[0\leq p_i, q_j, r_k \leq 1 \quad i,j,k=1,\dots, n.\]

The Fast NLP for three players is implemented in procedure \texttt{FastMNE(n,b)}, which returns one solution to mixed NE. Here, we assume that Players II and III adopt identical strategies. For example, with bet size 1, typing

\noindent {\tt M := FastMNE(4,1):} \\
\noindent {\tt [ListCF(M[1]),ListCF(M[2]),ListCF(M[3]),ListCF(M[4])];} \\
gives \\
\noindent {\tt [[0, 1/24, -1/48, -1/48], [2/3, 0, 0, 1], [0, 0, 1/4, 1], [0, 0, 1/4, 1]]}.

\vspace{1em}
Translation:
\begin{itemize}
    \item The value of the game (for Player 1) is $\frac{1}{24}$, while for Players II and III are $-\frac{1}{48}$ each.
    \item Player I's strategy is: If your card is 1, bet with probability of $\frac{2}{3}$ and check with probability $\frac{1}{3}$. If your card is 2 or 3, then definitely checks; if your card is 4, definitely bet.
    \item Player II's and III's strategies are: If their card is 1 or 2, they definitely fold. If their card is 3, they call with probability of $\frac{1}{4}$ and fold with probability $\frac{3}{4}$. If their card is 4, they definitely call. 
\end{itemize}

Another example, \\
\noindent {\tt M := FastMNE(10,2):} \\
\noindent {\tt[ListCF(M[1]),ListCF(M[2]),ListCF(M[3]),ListCF(M[4])];} \\
produces\\
\noindent {\tt[[0, 106/1125, -53/1125, -53/1125], [16/19, 0, 0, 0, 0, 0, 0, 0, 0, 1], [0, 0, 0, 0, 0, 0, 3/25, 1, 1, 1], [0, 0, 0, 0, 0, 0, 3/25, 1, 1, 1]]}, \\
which we leave for the reader to interpret. 

The verbose form of {\tt FastMNE(n,b);} is {\tt FastMNEVerbose(n,b);}, spelling out the advice. The output file {\small\url{https://sites.math.rutgers.edu/~zeilberg/tokhniot/oThreePersonPoker1.txt}} contains one mixed NE for each of the cases $n$ (size of the deck) from $5$ to $15$, and $b$ (size of the bet) from $1$ to $3$.

\subsection*{Extension of von Neumann's continuous model to three players}

As you may recall, we introduced this work with von Neumann's concept of an uncountably infinite deck, contrasting it with the finite nature of real-world card games. We proceeded by solving the finite deck games for two players and extended our analysis to include three players. Now that we have solved the finite version, the solutions effortlessly transition us to the continuous version of the three-player game, as we will demonstrate—a fitting conclusion to our study.

As in the von Neumann model for two players, each of the three players contributes 1 dollar to the pot and receives independent uniform(0,1) hands. As a reminder, Player I has the option to check or bet a fixed amount 
$b$, while Players II and III can only call or fold. The betting tree remains the same as  that of the finite deck model. The conjectured Nash equilibrium strategies for three players, guided by the data generated from the finite deck model, are illustrated in the right panel of Figure \ref{fig:Player3_strategies}.  Please scroll up a few pages.

We now determine the NE strategies. 

\subsubsection*{Our advice}

For numbers $A,B,C$, yet {\it to be determined},

\begin{itemize}
    \item Player I: If $0<x<A$ or $B<x<1$ he should {\bf bet}, otherwise {\bf check}.
    \item Players II and III: If $0<y< C$ they should {\bf fold}, otherwise {\bf call}.
\end{itemize}

Assume $0<A<C<B$.  To determine the cut points $A,B$ and $C$ we solve three indifference equations as follows. 

\begin{enumerate}
\item  \textbf{For Player I to be indifferent at $A$:}
\begin{enumerate}
    \item If Player I checks at $x=A$, his expected payoff is
    \[
    \intop_{0}^{A}\intop_{0}^{A} 2 dzdy+ \intop_{0}^{A}\intop_{A}^{1} -1 dzdy +\intop_{A}^{1}\intop_{0}^{1} -1 dzdy
    \]
    \item If Player I bets at $x=A$, his expected payoff is
        \[
    \intop_{0}^{C}\intop_{0}^{C} 2 dzdy+ \intop_{0}^{C}\intop_{C}^{1} -(b+1) dzdy +\intop_{C}^{1}\intop_{0}^{1} -(b+1) dzdy
    \]
\end{enumerate}
Equating the two expressions above yields the following equation:
\begin{equation}
3A^2-1=3C^2+bC^2-b-1.  \tag{Eq.~A}
\end{equation}

\item  \textbf{For Player I to be indifferent at $B$:}
\begin{enumerate}
    \item If Player I checks at $x=B$, his expected payoff is
    \[
    \intop_{0}^{A}\intop_{0}^{A} 2 dzdy+ \intop_{0}^{A}\intop_{A}^{1} -1 dzdy +\intop_{A}^{1}\intop_{0}^{1} -1 dzdy
    \]
    \item If Player I bets at $x=B$, his expected payoff is
        \begin{align*}
    &            \intop_{0}^{C}\intop_{0}^{C} 2 dzdy+ \intop_{0}^{C}\intop_{C}^{B} (b+2) dzdy +\intop_{C}^{B}\intop_{0}^{C} (b+2) dzdy
   +\intop_{C}^{B}\intop_{C}^{B} 2(b+1) dzdy\\
   & +\intop_{B}^{1}\intop_{0}^{B} -(b+1) dzdy
    +\intop_{0}^{B}\intop_{B}^{1} -(b+1) dzdy
    +\intop_{B}^{1}\intop_{B}^{1} -(b+1) dzdy
        \end{align*}
\end{enumerate}
Equating the two expressions above yields the following equation:
\begin{equation}
3B^2-1 = -2bCB+3bB^2+3B^2-b-1. \tag{Eq.~B}
\end{equation}

\item  \textbf{For Player II (or Player III) to be indifferent at $C$:}
\begin{enumerate}
    \item If Player II folds at $y=C$, her expected payoff is
    \[
    \intop_{0}^{A}\intop_{0}^{1} -1 dzdx+ \intop_{B}^{1}\intop_{0}^{1} -1 dzdx
    \]
    \item If Player II calls at $y=C$, her expected payoff is
        \[
    \intop_{0}^{A}\intop_{0}^{C} (b+2) dzdx+ \intop_{0}^{A}\intop_{C}^{1} -(b+1) dzdx +\intop_{B}^{1}\intop_{0}^{1} -(b+1) dzdx
    \]
\end{enumerate}
Equating the two expressions above yields the following equation:
\begin{equation}
-A + B - 1 = 2 b C A + 3 C A - b A - A - b + b B + B - 1. \tag{Eq.~C}
\end{equation}

\end{enumerate}

Solving the above {\bf non-linear} system of three equations in three unknowns gives us the solutions for $A,B$ and $C$ for the Nash equilibrium strategies. 
The procedure \texttt{Optimal(b);} returns these solutions, with the verbose version \texttt{OptimalVerbose(b);}. 

In particular, when $b=2$, \texttt{Optimal(2);} returns: 
\begin{align*}
A&=0.137058194328370\\
B&=0.829422249795391\\
C&=0.641304115985175.   
\end{align*}
This results in the value of the game (for Player I) being 0.122557074714865.

We can also determine the best bet amount $b$, that maximizes Player I's payoff under the Nash equilibrium strategies. Approximately, $b^*\approx 2.07$, resulting in Player I achieving a maximum payoff of 0.122590664136184. Therefore, we observe that the highest payoff for Player I in the three-player game exceeds that of the von Neumann's two-player game, which is 1/9 = 0.111111 achieved at $b^*=2$. 

One final remark is that the Nash equilibrium for the three-player continuous game resembles those observed in the discrete model when $n$ is large. In our experiments with the finite deck model, we can simulate up to $n=65$ in Maple, by executing \texttt{FastMNE(65,2);}, which yields: 

\noindent\texttt{[[0, 974/8121, -487/8121, -487/8121], [1, 1, 1, 1, 1, 1, 1, 1, 14/23, 0, 0, 0, 0, 0, 0, 0, 0, 0, 0, 0, 0, 0, 0, 0, 0, 0, 0, 0, 0, 0, 0, 0, 0, 0, 0, 0, 0, 0, 0, 0, 0, 0, 0, 0, 0, 0, 0, 0, 0, 0, 0, 0, 0, 0, 1, 1, 1, 1, 1, 1, 1, 1, 1, 1, 1], [0, 0, 0, 0, 0, 0, 0, 0, 0, 0, 0, 0, 0, 0, 0, 0, 0, 0, 0, 0, 0, 0, 0, 0, 0, 0, 0, 0, 0, 0, 0, 0, 0, 0, 0, 0, 0, 0, 0, 0, 0, 0, 189/205, 1, 1, 1, 1, 1, 1, 1, 1, 1, 1, 1, 1, 1, 1, 1, 1, 1, 1, 1, 1, 1, 1], [0, 0, 0, 0, 0, 0, 0, 0, 0, 0, 0, 0, 0, 0, 0, 0, 0, 0, 0, 0, 0, 0, 0, 0, 0, 0, 0, 0, 0, 0, 0, 0, 0, 0, 0, 0, 0, 0, 0, 0, 0, 0, 189/205, 1, 1, 1, 1, 1, 1, 1, 1, 1, 1, 1, 1, 1, 1, 1, 1, 1, 1, 1, 1, 1, 1]]} .

\vspace{1em}
Translation:
\begin{itemize}
    \item The value of the game (for Player I) is $\frac{974}{8121}=0.119935968476789$. 

    \item The cuts are: \quad $A= \left(8+14/23\right)/65=0.132441471571906$ \vspace{-0.5em}
    \begin{align*}
B&= 1-11/65=0.830769230769231 \\
C&= 1-\left(22+189/205\right)/65= 0.647354596622889.
\end{align*}
\end{itemize}

For more examples, please visit {\small\url{https://sites.math.rutgers.edu/~zeilberg/mamarim/mamarimhtml/poker.html}}  to explore the input-output files of the Maple programs implemented in this work.

\vspace{1em}
\noindent \hrulefill
\vspace{1em}

\begin{small}
\noindent Tipaluck Krityakierne, Department of Mathematics, Faculty of Science, Mahidol University,
272 Rama VI Rd., Ratchathewi,
Bangkok 10400, Thailand \hfil\break
Email: {\tt  tipaluck.kri at mahidol dot edu}

\vspace{1em}
\noindent Thotsaporn Aek Thanatipanonda, Science Division, Mahidol University International College, 
999 Phutthamonthon 4 Rd., Salaya, Phutthamonthon, Nakhon Pathom 73170, Thailand \hfil\break
Email: {\tt thotsaporn at gmail dot com}

\vspace{1em}
\noindent Doron Zeilberger, Department of Mathematics, Rutgers University (New Brunswick), Hill Center-Busch Campus, 110 Frelinghuysen
Rd., Piscataway, NJ 08854-8019, USA. \hfill\break
Email: {\tt DoronZeil at gmail  dot com} 
\end{small}
\vspace{1em}

\noindent {\bf July 22, 2024} 

\vfill

\centering
\includegraphics[scale=0.5]{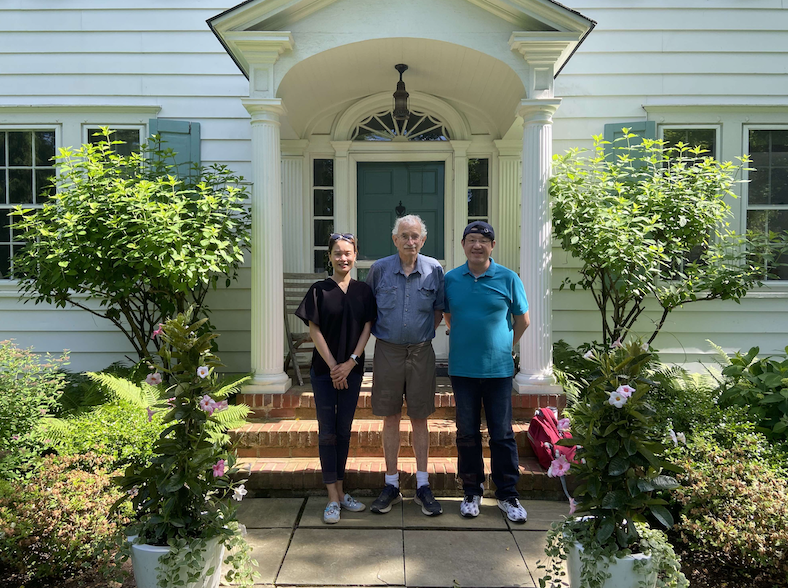}

\vspace{1em}
\textit{\small The authors are pictured in front of John von Neumann’s old house at 26 Westcott Road in Princeton, taken on May 31, 2024. The photo was taken by Karen Reid, the current owner, who kindly allowed us to take the photo.}

\end{document}